\documentclass[aps,prl,reprint,superscriptaddress,amssymb,amsmath]{revtex4-2}
\usepackage{graphicx}
\def\bra{\langle}
\def\ket{\rangle}
\def\<{\langle}
\def\>{\rangle}

\begin{document}
\title{Repeatable classical one-time-pad crypto-system  with quantum mechanics}
\author{Fu-Guo Deng}
\affiliation{Key Laboratory For Quantum Information and Measurements
and Department of Physics, Tsinghua University, Beijing 100084,
P. R. China}
\affiliation{ Key Laboratory of Beam Technology and Materials
Modification of MOE, and Institute of Low Energy Nuclear Physics,
Beijing Normal University, Beijing 100875, P. R. China}
\author{Gui Lu Long}
\affiliation{Key Laboratory For Quantum Information and Measurements
and Department of Physics, Tsinghua University, Beijing 100084,
P. R. China}
\affiliation{Center for Atomic and Molecular
NanoSciences, Tsinghua University, Beijing 100084, P. R. China}
\date{ Submitted to PRL as LU9745 on 29 July 2004}

\begin{abstract}
Classical one-time-pad key can only be used once. We show in this
Letter that with quantum mechanical information media classical
one-time-pad key can be repeatedly used. We propose a specific
realization using single photons. The reason why quantum mechanics
can make the classical one-time-pad key repeatable is that quantum
states can not be cloned and eavesdropping can be detected by the
legitimate users. This represents a significant difference between
classical cryptography and quantum cryptography  and provides a
new tool in designing quantum communication protocols and
flexibility in practical applications.\\

Note added: This work was submitted to PRL as LU9745 on 29 July 2004, and the decision was returned on 11 November 2004, which advised us to resubmit to some specialized journal, probably, PRA, after revision. We publish it here in memory of Prof. Fu-Guo Deng (1975.11.12-2019.1.18), from Beijing Normal University, who died on Jan 18, 2019 after two years heroic fight with pancreatic cancer. In this work, we designed a protocol to use a classical one-time-pad key of 2N length to prepare a sequence of N single photons in $|0>$, $|1>$, $|+>$, $|->$ states and encode secret message using unitary operations I (for 0), sigma-y (for 1). The bit string can be reused. The essential idea was proposed in November 1982, by Charles H. Bennett, Gilles Brassard, Seth Breidbart, which was rejected by Fifteenth Annual ACM Symposium on Theory of Computing, and remained unpublished until 2014, when they published the article, Quantum Cryptography II: How to re-use a one-time pad safely even if P=NP, Natural Computing (2014) 13:453-458, DOI 10.1007/s11047-014-9453-6. We worked out this idea independently. This work has not been published, and was in cooperated into quant-ph 0706.3791 (Kai Wen, Fu Guo Deng, Gui Lu Long, Secure Reusable Base-String in Quantum Key Distribution), and quant-ph 0711.1632 (Kai Wen, Fu-Guo Deng, Gui Lu Long, Reusable Vernam Cipher with Quantum Media).

\end{abstract}

%\maketitle
\maketitle

Quantum mechanics predictions can sometimes contradict our
experiences and our classical physics knowledge with surprises. It
is well-known that quantum mechanics allows for efficient solution
to difficult problems in classical computation, for instance in
Shor algorithm\cite{shor} and Grover's algorithm \cite{grover}. In
the fields of cryptography,  quantum key distribution(QKD)
provides unconditionally secure distribution of secret keys
between two remote parties\cite{bb84,chaulo,mayers,ShorBB84}.
Since the early QKD protocols \cite{bb84,ekert91}, research in QKD
has been progressing very fast\cite{gisin}.  The secure key
produced from QKD can be combined with the Vernam
one-time-pad\cite{vernam}, which has been proven unconditionally
secure\cite{shannon}. In the Vernam cipher scheme, the secret key
and the message have the same length, and the cipher text is the
simple modulo 2 sum of the message and the keys.

The Vernam one-time-pad key can only be used once in classical
cryptography. To an eavesdropper Eve, the message $M$ is unknown
to her, and the entropy of the message space is $H(M)$. Suppose
Eve gets hold of the ciphertext $C$ and her entropy about the $M$
becomes $H_C(M)$. With perfect secrecy, $H(M)=H_C(M)$, i.e., the
possession of the ciphertext does not provides Eve any new
information about the message. Vernam cipher is just such a
system. However when the same key is used twice, the perfect
secrecy condition is no longer satisfied. For instance if the
first message were obtained by Eve due to one reason or another,
she would have complete knowledge of the key by subtracting the
intercepted ciphertext with the message. Hence the repeated use of
the one-time-pad key in classical cryptography is terrifying and
there have been hard lessons in history of the disastrous
consequences of the repeated use of a one-time-pad key. However,
in this Letter, we will show that classical one-time-pad can be
used repeatedly with quantum mechanical carriers with
unconditional security. The essential cause for the repeated use
of the key is the inability of Eve to intercept the ciphertext and
the capability of communicating parties to detect Eve. The
significance of this work is twofold. First, the repeated use of a
classical one-time-pad key itself represents conceptual liberation
from the constraints laid down by classical cryptography. This may
provide new avenue for cryptography. Second, this property may
help present studies of quantum communications, both in protocol
design and in practical realizations.

First we briefly check the basic ingredients in a QKD protocol
that makes QKD secure. They are: 1) the quantum non-cloning
theorem\cite{nocloning}; 2) quantum state collapse after
measurement; 3) non-locality of entangled composite quantum
systems\cite{epr};(4) classical randomness. Usually a QKD protocol
involves simultaneously more than one of the ingredients mentioned
above. For example, the BB84 QKD protocol uses ingredient 2) and
4). The current practice in quantum communication is that first a
common key between  two users is established by a QKD, and then
the message is encrypted with the key using the Vernam cipher. The
ciphertext is transmitted from one user to the other user through
a classical channel. Recently, some authors have proposed to
transmit secret messages directly through a quantum
channel\cite{beige,bf,dl1,dl2}, which condenses the two
transmissions into a single one. In these schemes, the key is used
only once where in the former the ciphertext is transmitted
through a classical channel, whereas in the latter the ciphertext
is transmitted through a quantum channel.

Suppose Alice and Bob have already a sequence of common secret
key, and they have access to a quantum channel. Can they use the
key repeatedly? We will show with a explicit protocol that the
answer is yes.

\begin{figure}[tpb]
\begin{center}
\caption{ Illustration of a repeated classical one-time-pad
crypto-system. }
\includegraphics[width=8cm,angle=0]{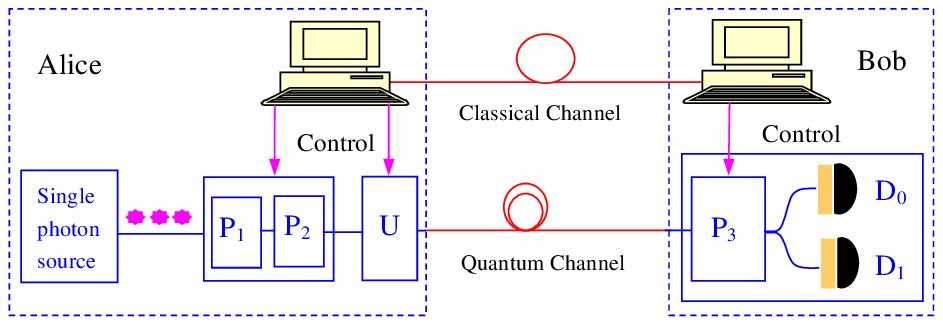} \label{f1}
\end{center}
\end{figure}

Like the BB84 QKD protocol, there are two sets of
measuring-basis(MB),  the plus-measuring-basis (plus-MB):
$$\left\vert H\right\rangle =\left\vert 0\right\rangle, \left\vert
V\right\rangle =\left\vert 1\right\rangle, $$ and the
cross-measuring-basis (cross-MB), i.e.
$$\left\vert u\right\rangle =\frac{1}{\sqrt{2}%
}(\left\vert 0\right\rangle +\left\vert 1\right\rangle ),
\left\vert d\right\rangle =\frac{1}{\sqrt{2}}(\left\vert
0\right\rangle -\left\vert 1\right\rangle ),$$
 where $|H\rangle $
and $|V\rangle $ are the horizontal and vertical polarization
states respectively. The quantum crypto-system contains  six
steps, the schematic illustration is shown in Fig.\ref{f1}. We
assume ideal noiseless channel. For completeness, we also include
the step to establish a common key sequence. We assume that Alice
is transmitting a message to Bob.

{\bf Step 1}: Establishing a classical one-time-pad key.  Alice
and Bob first establish a sequence of secret key. There are
various means to achieve this. One natural choice is  to use the
BB84 QKD protocol\cite{bb84} to produce a common secret key
between Alice and Bob. The result is a sequence of random binary
numbers.

{\bf Step 2}: Modified message sequence preparation. A secret
message can be translated into a sequence of $N_m$ binary numbers,
and the message sequence is denoted as $M_m$. The modified message
sequence contains the message sequence and a sampling sequence of
binary numbers. Alice chooses a sufficient large sequence of $N_s$
random binary numbers as sampling bits, denoted by $M_S$. Alice
inserts each of these $N_s$ binary numbers into the message bit
sequence at random positions. This comprises a modified message
sequence of $N=N_m+N_s$ binary numbers. This modified message
sequence is denoted as $M_N$. Alice records the position and the
value of these $N_s$ binary numbers in this modified message
sequence $M_N$.

{\bf Step3}: Encoding the modified message sequence with the
classical one-time-pad key. Take $2N=2(N_m+N_s)$ binary numbers
from the one-time-pad key to form a basis-key sequence, denoted by
$Q_N$ . Alice and Bob agree beforehand that  00, 11, 01 and 10 in
the classical one-time-pad base-key correspond to quantum states
$\left\vert H\right\rangle $, $\left\vert V\right\rangle $,
$\left\vert u\right\rangle $ and $\left\vert d\right\rangle $
respectively, namely \begin{eqnarray} 00\rightarrow |H\ket,
11\rightarrow |V\ket, 01\rightarrow |u\ket, 10\rightarrow |d\ket.
\end{eqnarray}

Alice prepares a sequence of $N$ single photons in states
according to the respective values in the basis-key $Q_N$. The
encoding of the modified message sequence is realized by two
unitary operations
\begin{eqnarray} U_0&=I&=|0\ket\bra 0|+|1\ket\bra 1|,\\
U_{1}&=i\sigma _{y}&=\left\vert 0\right\rangle \left\langle
1\right\vert -\left\vert 1\right\rangle \left\langle 0\right\vert.
\end{eqnarray}
If the i-th bit in the modified message sequence is a 0, then
Alice performs unitary operation $U_0$ on the i-th single photon,
and if the i-th bit in the modified message sequence is a 1, then
Alice performs  $U_1$ operation on the i-th single photon. The
nice feature of this set of encoding operation is they do not
change the states in one measuring-basis into another
measuring-basis,
\begin{eqnarray}
&U_{1}\left\vert 0\right\rangle =-\left\vert 1\right\rangle ,
&U_{1}\left\vert 1\right\rangle =\left\vert 0\right\rangle ,\nonumber\\
&U_{1}\left\vert u\right\rangle =\left\vert d\right\rangle ,\;\;
&U_{1}\left\vert d\right\rangle =-\left\vert u\right\rangle .
\label{U1}
\end{eqnarray}

After encoding the corresponding bit value in the modified message
sequence on the single photon, Alice sends the single photon to
Bob.  The single photons now carry the ciphertext.

{\bf Step 4}: Decrypting the ciphertext of the modified message
sequence by Bob. After receiving the single photons, Bob measures
each single photon in appropriate measuring-basis as Bob has the
same one-time-pad key as Alice so he knows the state of each
photon before the encoding. After the measurement, Bob knows the
encoding operation on each photon, and hence knows the
corresponding bit value in the modified message sequence $M_N$.

{\bf Step 5}: Alice and Bob check eavesdropping and Bob obtains
the message sequence. Alice announces publicly the positions of
the sampling bits in the modified message sequence. With this
information, Bob publicly announces the bits values of these
sampling bits, that is the encoding operations of the sampling
photons. It should be emphasized that the information about the
states of the photons are not disclosed, either before  the
encoding nor after the encoding. Alice then compares these results
with her own values, and she determines the error rate. With the
error rate, she can judge whether there is eavesdropping in the
transmission. Under ideal condition, there should be no error if
there is no eavesdropping. If the error rate is high, then there
is eavesdropper in the transmission. They halt the process.

{\bf Step 6}: Constructing a classical one-time-pad key from the
used classical one-time-pad key. Under ideal condition, if the
error rate is zero, then Alice and Bob can  drop the sampling bits
that have been chosen for eavesdropping and keep the remaining
bits in the classical one-time-pad keys as a new classical
one-time-pad key to be used later.

We now discuss the security of this reusable classical
one-time-pad key, $Q_{k}$. We do not prove the security here, but
rather point only to the similarity with the classical one-time
pad \cite{vernam} and BB84 QKD \cite{bb84} which are known to be
unconditionally secure\cite{shannon,chaulo,mayers,ShorBB84}.

First we notice that the encoding of secret messages in the step 3
is identical to the process in a classical one-time-pad where the
text is encrypted with a random key as the states of the photons
in the quantum key is completely random to Eve, hence the
conditional entropy of the message $H_c(M)$ is identical to the
entropy of the message itself $H(M)$, $H_c(M)=H(M)$, i.e., the
access of the ciphertext by Eve does not increase any information
about the message. This is completely secure according to Shannon
\cite{shannon}. However, as has been discussed in Ref. \cite{dl2},
the quantum crypto-system, is even more secure than the classical
one-time-pad scheme in the sense that Eve can not even get the
cipher-text completely, as she does not know the photons' MBs, and
any eavesdropping will 1) destroys the quantum state; 2) be
detected by Alice and Bob. Thus the security of this quantum
crypto-system depends entirely on the security of the quantum key.
If this one-time-pad key were lost to Eve, she could simply
measure each of the single photon with the correct measuring-basis
and escape detection.   We will discuss the security issue of this
quantum crypto-system. We limit our discussion to individual
attack first \cite{WZY,fggnp,NC} and discuss the security of
$Q_{k}$ .

In the ideal condition, any eavesdropping done by Eve will be
detected by Alice and Bob. Moreover, Eve can get nothing about the
ciphertext. The process for eavesdropping check in this quantum
crypto-system is similar as that for BB84 QKD protocol\cite{bb84}
in which the optimal individual attack of Eve  can be realized by
an unitary operation $U_{TB}$ on the travelling photon sent to Bob
with a probe whose initial state is $\left\vert 0\right\rangle $,
i.e.,
\begin{eqnarray}
U_{TB}\left\vert \xi \right\rangle \left\vert 0\right\rangle
&=&\left\vert \xi \right\rangle \left\vert 0\right\rangle,
\label{ut1}\\
U_{TB}\left\vert \overline{\xi }\right\rangle \left\vert
0\right\rangle &=&\cos \theta \left\vert \overline{\xi
}\right\rangle \left\vert 0\right\rangle +\sin \theta \left\vert
\xi \right\rangle \left\vert 1\right\rangle  \label{ut2}
\end{eqnarray}%
where $\left\vert \xi \right\rangle $ and $\left\vert \overline{\xi }%
\right\rangle $\ are two eigenvectors of two-level operator, such
as $\sigma _{z}$ or $\sigma _{x}$, and $\theta \in \lbrack 0,\pi
/4]$ represents the strength of the eavesdropping
\cite{fggnp,ssz,sg}. The average mutual information $I_{AE}$
between Eve and Alice is limited to\cite{fggnp},
\begin{equation}
I_{AE}\leq \frac{1}{2}\phi \left[ 2\sqrt{D(1-D)}\right] \equiv
I_{0} \label{information1}
\end{equation}%
where the function $\phi \left[ x\right] =(1+x)\log
_{2}(1+x)+(1-x)\log _{2}(1-x)$, and $D=\frac{1}{2}\sin ^{2}\theta
$\ is the error rate introduced by Eve's action in the result.

It is useful to emphasize that $I_{0}$ is the information bound
Eve can obtain about the encoding operation sent by Alice under
the condition that Alice and Bob publish the information about the
measuring-basis.  When $D=0.25$,  $I_{0}\approx 0.645$ is the
maximum.

In the protocol described above, Alice does not publish any
information of the single photons of the message, and Eve's
information $I_{AE}^{^{\prime }}$ is less than $I_{0}$
\begin{equation}
I_{AE}^{^{\prime }} < I_{AE}\leq I_{0} < 1 = I_{AB},
\label{information2}
\end{equation}%
where $I_{AB}$ is the mutual information between Alice and Bob. It
is obvious that Eve cannot get a useful information about the
message and her action will be detected.

When producing the one-time-pad from the used one, Alice and Bob
discard $M_S$ sequence which has been publicly announced. They use
the one-time pad repeatedly only when they ascertain the quantum
channel is secure. So Eve can get nothing about the message
without information about it.

A special case is that Eve eavesdrops the quantum channel with
some information about the secret message known in advance. The
goal is to obtain some information about the classical one-time
key. It is of interest that even though Eve knows the message
before hand and attacks the quantum communication with cipher-text
strategy, she still cannot get the one-time-pad key completely as
she cannot get the information about the quantum states of single
photons without their MBs. This is certainly different from that
in classical cryptography. In this time, the information
$I_{AE}^{^{\prime\prime }}$ is just what Eve can get about the
basis-key of the single photon when she knows the encoding
operation in advance.

Similar to the BB84 QKD scheme in which Eve wants to obtain some
information about the raw key and monitors the quantum
channel\cite{NL}, Eve eavesdrops the quantum communication for the
classical one-time pad key. The mutual information
$I_{AE}^{^{\prime \prime}}$ is just the information that Eve can
get about the measuring-basis of the single photons. Eve can get
more information about the raw key in BB84 QKD than that in this
crypto-system for the MBs of the single photons as she can do
exact measurement on her probe system and get the result with the
MBs published by Alice and Bob in BB84 QKD, which does not happen
in this crypto-system. So we can get the relation\cite{NL}
\begin{equation}
I_{AE}^{^{\prime \prime}}<I_{AEBB84} \leq 1-\log _{2}(1+\widetilde{\varepsilon })+\frac{%
\widetilde{\varepsilon }}{1+\widetilde{\varepsilon }}\log _{2}\widetilde{%
\varepsilon }\equiv I_{1},  \label{information3}
\end{equation}%
\begin{equation}
\widetilde{\varepsilon }\geq \left( \frac{1-4\sqrt{\left( \sqrt{2-8D_{m}}%
\right) D_{m}}}{1-8\sqrt{2}D_{m}}\right) ^{2}  \label{min}
\end{equation}%
where $I_{AEBB84}$ is mutual information between Alice and Eve in
BB84 QKD, the maximum value is $D_{m}\leq 0.25$. As pointed out in
Ref. \cite{NL},
 for small $D_{m}$,%
\begin{equation}
I_{AE}^{^{\prime \prime}}< \frac{4\sqrt{2}}{\ln 2}D_{m},
\label{information4}
\end{equation}%
where ln2 is the natural logarithm of 2. For example, when $D_{m}\leq 0.05$, $%
I_{AE}^{^{\prime \prime }}< 0.41$. As discussed above, Eve cannot
steals the secret message if she only obtains the information
$I_{MB}$, but it is helpful for her to eavesdrop the quantum
signal in the next time. This is just the reason that our quantum
crypto-system can repeatedly use the quantum key conditionally.

In ideal implementation of this quantum crypto-system, single
photon source is required. A single photon source is principally
available\cite{single}. Even without single photon source, ideal
channels ad perfect detectors, this repeatable one-time-pad scheme
can also be used. Error correcting techniques are necessary then.
There have already been quite a few good correcting codes, for
instance, in references\cite{cascade,css,feng}. Comparing with the
quantum one-time-pad secure direct communication scheme in
Ref.\cite{dl2}, the advantage of this protocol is obvious: there
is no need to store quantum states, and the single photons need be
sent only in one-way, as compared with that in Ref.\cite{dl2} with
two-ways.

With a noise quantum channel, this crypto-system can be used to
quantum direct communication with some other quantum and classical
techniques if the noise is low. For example Alice and Bob do
privacy amplification \cite{pa} on the classical one-time-pad key
for cancelling the information of the key leaked to Eve. The
process is just a map in which $k$ bits of key is changed to
$k^{\prime}$, where $k>k^{\prime}$. Surely, this crypto-system
maybe not suitable for direct communication if the noise and loss
are high as there are no advantages compared with QKD. In the
condition, it is suitable for QKD similar to the ways in Refs
\cite{HKH,DL}.

To summarize, this quantum crypto-system can be used for
transmission of secret message securely, same as QSDC even though
it is just a prototype. Also, it can be used for QKD
unconditionally securely. In this quantum crypto-system, the
quantum key, a sequence of the quantum states of single photons
can be used repeatedly and is mapped to a classical key which is
stored in classical way.

This work is supported the National Fundamental Research Program
Grant No. 001CB309308, China National Natural Science Foundation
Grant No. 60073009, 10325521, the Hang-Tian Science Fund, the
SRFDP program of Education Ministry of China.

\end{document}